\documentclass[aps,preprint,nofootinbib,floatfix]{revtex4-1}
\usepackage{color}

\pdfoutput=1
\usepackage{graphicx}
\usepackage[latin1]{inputenc}
\usepackage{amsmath,amssymb}
\usepackage{hyperref}
\usepackage{epstopdf}

\newcommand{\lsim}{\mathrel{\mathop{\kern 0pt \rlap
  {\raise.2ex\hbox{$<$}}}
  \lower.9ex\hbox{\kern-.190em $\sim$}}}
\newcommand{\gsim}{\mathrel{\mathop{\kern 0pt \rlap
  {\raise.2ex\hbox{$>$}}}
  \lower.9ex\hbox{\kern-.190em $\sim$}}}

\newcommand{\eps}{\varepsilon}

\def  \bcen   {\begin{center}}
\def  \ecen   {\end{center}}
\def  \beq    {\begin{equation}}
\def  \eeq    {\end{equation}}
\def  \bpm    {\begin{pmatrix}}
\def  \epm    {\end{pmatrix}}
\def  \beqa   {\begin{eqnarray}}
\def  \eeqa   {\end{eqnarray}}
\def  \nn     {\nonumber }
\def\bea{\begin{eqnarray}}
\def\eea{\end{eqnarray}}

\def\ga   {\gamma}
\def\Ga   {\Gamma}
\def\th   {\theta}

\def\la   {\lambda}
\def\La   {\Lambda}
\def\al   {\alpha}

\def\nn{\nonumber}
\def\lee { \left( }
\def\rii { \right) }
\def\lan   {\langle}
\def\ran   {\rangle}
\def\de {\delta}
\def\De {\Delta}
\def\ka  {\kappa}

\def\to {\rightarrow}
\def\vep {\varepsilon}

\begin{document}
 
 {\small
 \begin{flushright}
 DO-TH 16/22 \\
 \end{flushright} }

\linespread{1.2}

\title{ \boldmath Sizable  NSI  from the $SU(2)_L$ scalar doublet-singlet mixing and the implications in DUNE}

\author{David V. Forero}\email{dvanegas@vt.edu}
\affiliation{Center for Neutrino Physics,  Virginia Tech, Blacksburg, VA 24061, USA}
\author{Wei-Chih Huang}\email{wei-chih.huang@tu-dortmund.de}
\affiliation{Fakult\"at f\"ur Physik, Technische Universit\"at Dortmund,
44221 Dortmund, Germany}


\begin{abstract}
We propose a novel and simple mechanism where sizable effects of non-standard interactions~(NSI) in neutrino propagation
are induced from the mixing between an electrophilic second Higgs doublet and a charged singlet.
The mixing arises from a dimensionful coupling of the scalar doublet and singlet to the standard model Higgs boson.
In light of the small mass, the light mass eigenstate from the doublet-singlet mixing can generate much larger NSI than those
induced by the heavy eigenstate.
We show that a sizable NSI $\varepsilon_{e\tau}$~($\sim 0.3$) can be attained without being excluded by a variety of experimental
constraints. Furthermore, we demonstrate that NSI can mimic  effects of the Dirac CP phase in the neutrino mixing matrix but they
can potentially be disentangled by future long-baseline neutrino experiments, such as the Deep Underground Neutrino Experiment (DUNE).  
\end{abstract}

\maketitle
\bigskip
\section{Introduction and general motivations
\label{section:intro}}

Neutrino oscillations (in the three neutrino framework) are the leading 
mechanism that explain neutrino flavor transitions observed from neutrinos produced in the 
sun, the Earth atmosphere, reactors, and accelerators. The parameters accounting for 
neutrino oscillations, the three mixing angles and the two mass squared differences, have 
been currently measured within a precision of $8\%$ according to a global fit 
analysis~\cite{Forero:2014bxa}. One additional parameter, which encodes the violation of 
the charged parity (CP) symmetry in the lepton sector, is still to be determined. This 
parameter together with the determination of the neutrino mass ordering (normal or 
inverted hierarchy) are the two main unknowns in the three neutrino framework. 
Current and future facilities are aimed to find the two missing pieces and to improve the 
precision of the oscillation parameters. 

Better understanding of uncertainties on both theory and experiment sides is 
crucial in the completion and improvement of our knowledge of the three active neutrino 
framework. The reactor mixing angle has been measured within a precision of $5\%$ by $\sim 
1\text{km}$ baseline reactor neutrino multidetector experiments 
\cite{An:2015rpe,RENO:2015ksa}. The measurement of the atmospheric parameters (the mixing 
angle and the mass squared difference) have also been improved in precision thanks to the observations 
in the disappearance channel by beam-based neutrinos experiments, but still the atmospheric 
mixing angle is less well-determined among the three mixing angles. However, 
thanks to the tension in the determination of the reactor mixing angle by current reactor 
and accelerator experiments, an indication of preferred values for the Dirac CP violating phase have 
started to emerge~\cite{Abe:2015awa,Adamson:2016tbq}. This has opened the possibility of 
observing CP violation in the lepton sector, which might have an impact in the early Universe. 
Despite all success so far, upgrades of current and new facilities are needed to 
probe most of the Dirac CP parameter space, to determine the neutrino mass hierarchy and 
to improve the precision of the other parameters.

In addition to the standard three neutrino oscillation framework, there are well-motivated scenarios
beyond the standard model~(SM) that can have phenomenological consequences 
in neutrino oscillations. This opens the possibility to test new physics along
with the standard programs pursued in neutrino oscillation facilities. It would be, for 
instance, interesting to investigate non-standard neutrino interactions (NSI), 
non-unitarity neutrino mixing, sterile neutrinos, violation of symmetries, etc.
NSI were originally proposed even before neutrino oscillations were 
proved~\cite{Wolfenstein:1977ue,Valle:1987gv,Roulet:1991sm,Guzzo:1991hi} and still 
today their phenomenological consequences are being studied. NSI can be a byproduct of 
neutrino mass models and in general are described by effective four-fermion interaction operators 
where the strength is characterized by dimensionless couplings (carrying all the flavor 
information) times the Fermi constant. NSI can be of the charged-current (CC) type or 
of the neutral-current type (NC) depending on the fields involved,
and both of types have distinctive phenomenological consequences. 
The NSI yield additional contributions to the SM weak interactions and therefore
constraints can be derived, for instance, from 
lepton universality and CKM unitarity~\cite{Biggio:2009nt}. In cases where new physics enter above the
electroweak scale, both the 
charged and the neutral sectors are connected (due to the $SU(2)_L$ symmetry) and thus 
stringent constraints from charged lepton flavor violating~(CLFV) processes can have an impact 
on the neutral sector~\cite{Bergmann:1999pk}. 
The highlight of this work is to provide a simple mechanism to obtain large NSI effects and simultaneously
avoid these constraints such that one can determine the NSI strength via 
neutrino oscillation experiments. There exist many 
particular examples in the literature. In general, the CC-like NSI affects 
neutrino production and detection and can be cleanly probed in experiments where 
neutrino--matter interactions can be neglected, as in reactor neutrino experiments~ 
\cite{Agarwalla:2014bsa} (see also Ref.~\cite{Ohlsson:2013nna}) \footnote{Both CC and NC NSI 
can be tested at the same time in a long-baseline experiment, however, the large number 
of parameters will decrease the sensitivity for some NSI couplings.}. The NC-like NSI 
affects the neutrino propagation and can be probed in long-baseline neutrino oscillation 
experiments since the sensitivity is driven by the neutrino--mater interactions. The 
current NSI constraints, considering neutrino oscillations only, can be found 
in Ref.~\cite{Gonzalez-Garcia:2013usa}. For a general review of the NC-like NSI 
constraints and phenomenological implications, we refer the reader to 
Ref.~\cite{Miranda:2015dra} and references therein.

Among future experiments, the Deep Underground Neutrino Experiment (DUNE) is the main 
project that will determine the neutrino mass ordering and probe most parameter space
of the Dirac CP violating phase. DUNE will use a powerful beam to produce a large number 
of neutrinos in a broad energy range (roughly between 0.5 and 20 GeV) that 
will be detected in a $40\text{t}$ far detector located at $1300\text{km}$ from the 
source~\cite{Acciarri:2015uup}. As a result, DUNE will be an 
interesting NSI laboratory. This has been the subject of different studies 
showing that DUNE will be sensitive to the NC-like NSI with couplings of the order of 
$0.1 G_F$ (see 
for instance~\cite{Masud:2015xva,deGouvea:2015ndi,Coloma:2015kiu}). More importantly, 
degeneracies between the NSI couplings and the standard oscillation parameters might 
challenge the precise determination of the unknown neutrino parameters. This 
is the case for the determination of the Dirac CP violating phase; the NSI new phases 
are a new 
source of the CP violation and one might observe CP violation effects, which result 
exclusively from NSI. In a minimal setup, it has been shown that `confusion' 
can arise with an NSI parameter ($\eps_{e \tau}\sim 0.3$) in T2K and 
NOvA~\cite{Forero:2016cmb,Liao:2016hsa} (see also Ref.~\cite{Friedland:2012tq}
in which the `confusion' from $\varepsilon_{e\tau}$ was examined, after the measurement of
$\th_{13}$, at the probability level)~\footnote{For 
an analytic study of the CP determination in the presence of NSI at low energies, see 
Ref.~\cite{Ge:2016dlx}. }.

From the model building point of view, however, it is very challenging to come up with 
viable models which can produce such `large' NSI couplings and  avoid the constraints from 
 CLFV processes. As a result, the main goal of this work is 
to introduce a mechanism that generates relative large NSI couplings~($\sim 0.3$)
such that the aforementioned confusion can be realized. Future long-baseline neutrino 
oscillation experiments such as DUNE, can potentially resolve the confusion
and investigate the phenomenological implications of the NSI.

We propose a novel and simple mechanism to obtain large NSI $\vep_{e\tau}$.
In addition to the SM, there exist an extra $SU(2)$ scalar doublet  $\eta$ and a charged $SU(2)$ scalar singlet $\phi$. The pertinent scalar potential, including the SM Higgs doublet $H$,
is
\begin{align}
V \supset
 \mu_\eta^2 \eta^\dag \eta+\mu^2_\phi \phi^+ \phi^- + \lee \ka\, \phi^-  H  \eta  + h.c. \rii \, .
\end{align} 
The mixing between $\phi$ and the charged component of $\eta$ arises due to the coupling to the SM Higgs boson, $\ka\, \phi^-  \lan H \ran \eta$,
where $\ka$ is a dimensionful coupling and $\lan H \ran=v$.
In the limit of $\mu^2_\phi , \ka v \ll \mu^2_\eta $, the $\eta-\phi$ mixing is determined by the ratio of $\ka v$ to $\mu^2_\eta$
while the mass of the light eigenstate $m_1$ is determined by $\mu_\phi$ and the $\ka v / \mu_\eta $. 
These two components can cancel each other such that $m_1$ can be treated as an independent parameter from the $\eta-\phi$ mixing angle.
The independence is pivotal to achieve large NSI, satisfying various bounds from charged lepton measurements.  
  
Yukawa couplings of $\eta$ to leptons are introduced, obeying an imposed $Z_2$ symmetry, under which $\eta$, $\phi$ and the right-handed electron
$e_R$ are odd. NSI can be generated through charged currents mediated by the charged component $\eta^\pm$, which is a superposition of the two mass eigenstates   
in light of the $\eta-\phi$ mixing. The light mass eigenstate contribution to NSI can have a large enhancement due to its small mass even if it is suppressed by the mixing angle,
in that the mass $m_1$ is independent of the mixing. In other words, the mixing effect induces an additional contribution from the light eigenstate
which can be much larger than the heavy state contribution. Furthermore, large NSI realized via cancellation require fine-tuning and as demonstrated below,
taking into account various constraints, the level of $10^{-3}$ fine-tuning is needed for have sizable $\vep_{e \tau}$~($\sim 0.3$).

NSI mediated by $\eta$ alone are classified as dimension-6~($d-6$) operators in Ref~\cite{Gavela:2008ra} which usually comes with hazardous contributions to CLFV processes,
while the mixing-induced NSI belong to $d-8$ operators~(in light of extra $\lan H \ran^2$ compared to the $d-6$ one) and is usually suppressed with respective to $d-6$ ones.
As pointed out in Refs.~\cite{Berezhiani:2001rs,Davidson:2003ha}, some of $d-8$ operators only induce lepton flavor violation on the neutral sector but not on
the charged lepton counterpart such that stringent constraints on charged lepton flavor violation can be escaped.  
In our model, for instance, $\tau \to 3 e$ can be engendered by 
$\tau^- \to e^- \eta^0 \to e^- e^+ e^-$
but not mediated by $\phi^\pm$~(responsible for NSI). Consequently, sizable NSI do {\it not} imply large CLFV effects.  
Furthermore, by virtue of the cancellation within $m_1$, the mixing-induced
$d=8$ operators become dominant over those of $d=6$.

On the other hand, with the charged singlet $\phi$, the neutrino mass can be produced by adding an interaction $\overline{L^c} L \phi^+$, as proposed by Zee~\cite{Zee:1980ai,Zee:1985rj}.
The interaction itself can also yield NSI but it has been shown~\cite{Ohlsson:2009vk} that considerable NSI will demand large couplings of $\overline{L^c} L \phi^+$, rendering neutrinos too heavy.
In contrast, our mechanism is based on cancellation to enhance NSI $\vep_{e \tau}$ without involving the lepton number violating term $\overline{L^c} L \phi^+$, which is actually forbidden
by the imposed $Z_2$ symmetry.
Finally, models with a light gauge boson $Z^\prime$~\cite{Farzan:2015doa,Farzan:2015hkd,Machado:2016fwi} have been proposed
to generate considerable NSI but due to various bounds,
$\vep_{e \tau}$ is constrained to be much less than $0.3$.

The paper is organized in the following. In Section~\ref{section:model}, we
specify the model setup, followed by discussion of how sizable NSI can be attained via 
the doublet-singlet mixing in Section~\ref{section:NSI}. Various constraints are taken 
into account in Section~\ref{section:constraints}. Then we perform the numerical analysis 
in the context of long-baseline neutrino experiments in Section~\ref{section:analysis}.
Finally, we conclude in Section~\ref{section:conclusions}.

\section{Model \label{section:model}}
We enlarge the SM particle content by including two scalar fields, one $SU(2)_L$ doublet $\eta$ and one charged singlet $\phi$.
Furthermore, we impose a $Z_2$ symmetry under which $\eta$, $\phi$ and the right-handed electron are odd while the rest of SM particles are even: 
\begin{equation}
\eta \sim (\mathbf{1},\mathbf{2},+\tfrac{1}{2}, -)\;\; , \;\;
\phi^-  \sim (\mathbf{1},\mathbf{1},-1, -) \;\; , \;\; 
e_R \sim (\mathbf{1},\mathbf{1},-1, -),
\nonumber
\end{equation} 
where the entries in the parentheses denote the SM $SU(3)_c \times SU(2)_L \times U(1)_Y$ quantum numbers as well as
the $Z_2$ parity.
The reason of including the $Z_2$ symmetry is to avoid a myriad of experimental constraints from the charged lepton sector.
Note that the $Z_2$ symmetry is broken by the SM electron Yukawa coupling and
it is arguable that the smallness of the coupling results from the $Z_2$ symmetry protection.  

The relevant terms in the scalar potential read,
\begin{align}
V \supset
 \mu_\eta^2 \eta^\dag \eta+\mu^2_\phi \phi^+ \phi^- + \lee \ka\, \phi^-  H  \eta  + h.c. \rii \, ,
\end{align} 
where $H$ is the SM Higgs doublet, $\ka$ is a dimensionful coupling and $\mu^2_{\phi,\eta} >0$.
Note that we focus on regions of the parameter space where $\phi$ and $\eta$ do not develop the vacuum expectation value~(VEV).

 On the other hand, the mixing between $\phi^\pm$ and $\eta^{\pm}$ arises due to the SM Higgs VEV $v$,
and the mass matrix of $\phi^\pm$ and $\eta^\pm$ is given by
\begin{align}
M^2_{\eta \phi}=
  \begin{pmatrix}
    \mu_\phi^2   &    \ka v   \\[3mm]
      \ka v  &   \mu_\eta^2  \\
  \end{pmatrix} .
\end{align}
In the limit of  $\mu_\phi\,,\sqrt{\ka v} \ll \mu_\eta$, the masses of the two eigenstates $s_1$ and $s_2$ are
\begin{align}
m_1^2& \simeq\mu^2_{\phi} -   \theta\ka v + \mathcal{O}(\theta^2)\nn \\
m^2_2& \simeq\mu^2_\eta + \mathcal{O}(\theta)
\end{align}
with $\theta\simeq \ka v/\mu_\eta^2$.
Note that because of cancellation between $ \mu^2_{\phi}$ and $ \theta\ka v (= \ka^2 v^2/ \mu^2_\eta)$, $m_1$ can be treated as an independent parameter
from the mixing angle although, without any fine-tuning, it is expected that $m_1^2 \sim \mu^2_\phi \sim \ka^2 v^2/ \mu^2_\eta$ and $\th \sim m_1/m_2$.

Finally, we would like to point out that the neutrino mass can be generated
by adding $\overline{L^c} \phi L^c$~(Zee model~\cite{Zee:1980ai,Zee:1985rj}),
which however breaks the $Z_2$ symmetry, or simply by including heavy right-handed neutrinos~(Type-I seesaw).  The correlation between the neutrino mass mechanism and NSI, however, will not be explored here.

\section{NSI}\label{section:NSI}

To realize NSI, we couple the $SU(2)_L$ doublet $\eta$ to SM leptons via a renormalizable operator.
In light of $Z_2$  under which $\eta$, $\phi$ and the right-handed electron $e_R$ are odd, the only allowed term is
\begin{align}
\mathcal{L} \supset{\la_{\al}}\,  \overline{L_\al}  \, \eta \,  e_{R} + h.c. ,
\label{eq:eta_Yuk}
\end{align}
where $ \al$ is the flavor index, representing $e$ and $\tau$ but not $\mu$ in that
we concentrate on effects of $\vep_{e \tau}$, relevant for the confusion
mentioned above.

Neglecting the $\eta-\phi$ mixing, the effective operator of the charged current, after integrating out heavy $\eta$, reads
\begin{align}
\De \mathcal{L} \supset  \frac{\la_\al \la^*_\beta}{m^2_\eta} \lee \bar{\nu}_\al e_R \rii \lee \overline{e_R} \nu_\beta \rii
= \frac{\la_\al \la^*_\beta}{2 m^2_\eta}  \lee \bar{\nu}_\al  \gamma^\mu \nu_\beta \rii  \lee  \overline{e_R} \ga_\mu  e_R \rii ,
\end{align}
where the second equality comes from Fierz transformation.
Comparing with the charged current mediated by the $W$ boson,
\begin{align}
\De \mathcal{L} \supset - 2 \sqrt{2} G_F \lee \bar{\nu}_e \ga^\mu \nu_e  \rii \lee \overline{e_L} \ga_\mu e_L  \rii ,
\end{align}
one can obtain 
\begin{align}
\vep^\eta_{\al\beta}= - \frac{ \la_\al \la^*_\beta}{ 4 \sqrt{2} \, m^2_\eta G_F}.
\label{eq:eps_al_be}
\end{align}
After taking into account the $\eta-\phi$ mixing, the NSI from the two mass eigenstates $s_1$ and $s_2$ are
\begin{align}
\vep^{s_1}_{\al\beta} &= - \frac{ \la_\al \la^*_\beta \theta^2}{ 4 \sqrt{2} \, m^2_1 G_F} \nn \\
\vep^{s_2}_{\al\beta} &= - \frac{ \la_\al \la^*_\beta}{ 4 \sqrt{2} \, m^2_2 G_F} \, .
\label{eq:s12_NSI}
\end{align}  
 
Now, we can estimate the magnitude of the NSI from $s_1$ and $s_2$.
The $s_2$-induced contribution, assuming $\la_e \sim \la_\tau  \sim \la$, is
\begin{align}
\vep^{s_2}_{\al\beta} = - 3.9 \times 10^{-4} \lee \frac{  \la}{ 0.16} \rii^2 \lee \frac{\text{TeV}}{m_2}\rii^2 ,
\end{align}  
 while the $s_1$ contribution can be rewritten as
 \begin{align}
\vep^{s_1}_{\al\beta} = - 3.9 \times 10^{-4} \lee \frac{\th}{ m_1/m_2 } \rii^2 \lee \frac{  \la}{ 0.16} \rii^2 \lee \frac{\text{TeV}}{m_2}\rii^2.
\label{eq:NSI_s1e}
\end{align}  
 
 As we shall see below, due to the constraint on the CLFV process $\tau \to 3e$, $\la$ is restricted to be smaller than 0.16
 for TeV $s_2$. It implies the $s_2$-induced NSI contribution can not be considerable.
 Nevertheless, the $s_1$ contribution can be large since $m_1$ and $\th$ can be regarded as independent, i.e.
 $m_1/m_2$ can be quite different from $\th$. To realize a sizable NSI, one must have
 \begin{align}
 \frac{m^2_1}{m^2_2} \lesssim 10^{-3} \theta^2   \;\; \Rightarrow \;\;   \frac{\mu^2_\phi - \ka^2 v^2/ \mu^2_\eta}{ \frac{\ka^2 v^2}{\mu^2_\eta} } \lesssim 10^{-3} .
 \label{eq:FT_le}
  \end{align}
  It implies that in order to obtain a sizable NSI contribution of order $\mathcal{O}(0.1)$, the fine-tuning on the cancellation between $\mu^2_\phi$ and $\ka^2 v^2/ \mu^2_\eta$
  is required to be around the level of $0.1\%$.

\section{Constraints \label{section:constraints}}
Due to the existence of the couplings of $\phi$ and $\eta$ to the SM leptons, we here consider constraints involving charged leptons $e$ and $\tau$
from various measurements.

\begin{itemize}

\item LEP constraints on the mass of $s_1$.\\
 From the LEP measurements on the $Z$ decay width, the non-SM contribution are bounded below $2.9$ MeV~\cite{ALEPH:2005ab}, which requires that  
$m_1$ should be larger than half of the $Z$ mass to kinetically forbid $Z$ decay into $s_1^+ s^-_1$.
Besides, the LEP charged Higgs~($H^\pm$) searches~\cite{Abbiendi:2013hk} based on $e^+ e^- \to Z \to H^+ H^-$,
followed by $H^\pm \to \tau^\pm \nu$, set a limit of $m_{H^\pm}>80$ GeV in the context of two Higgs doublet models.
It also applies to $s_1$ in our model. Therefore, we have $m_1 \gtrsim 80$ GeV.

\item LEP $e^+ e^- \to \ell^+ \ell^-$ constraints. \\
The LEP measurements on the cross-section of $e^+ e^- \to \ell^+ \ell^-$ can be translated into
constraints on the new physics scale in the context of effective four-fermion interactions~\cite{LEP:2003aa}
\begin{align}
\mathcal{L}_{\text{eff}}= \frac{4\pi}{ \lee1 + \de \rii \Lambda^2} \sum_{i,j=L,R} \eta_{i,j} \bar{e}_i \ga_\mu e_i \bar{f}_j \ga^\mu f_j \, ,
\end{align}
where $\de=0~(1)$ for $f\neq e~(f=e)$ and $\eta_{ij}=1~(-1)$ corresponds to constructive~(destructive) interference between the SM and  new physics processes. 

In our model, $e^+ e^- \to \ell^+ \ell^-$ processes will be mediated by solely $\eta^0$ of mass $m_\eta~(\simeq m_2)$, which can be 
described by effective operators  
\begin{align}
\mathcal{L}_{\text{eff}}= \frac{\vert \la_e \vert^2 }{2 m^2_2}  \lee \overline{e_L}  \gamma^\mu e_L \rii  \lee  \overline{e_R} \ga_\mu  e_R \rii 
+  \frac{\vert \la_\tau \vert^2 }{2 m^2_2}   \lee  \overline{e_R} \ga_\mu  e_R \rii \lee \overline{\tau_L}  \gamma^\mu \tau_L \rii  .
\end{align}
Since $\La=9.1$ TeV for $e^+ e^- \to e^+ e^-$ and $\La=10.2$ TeV for $e^+ e^- \to \tau^+  \tau^-$~\cite{LEP:2003aa},
we infer $\la_e/m_2  \lesssim 0.39/\text{TeV}$  and $\la_\tau/m_2 \lesssim  0.49/ \text{TeV}$.

\item LEP mono-photon constraints. \\
Finally, the last LEP constraint comes from  DM searches based on the mono-photon signal~\cite{Fox:2011fx}:
$e^+ e^- \to \text{DM DM} \, \ga$ where $\ga$ comes from the initial state radiation or the internal bremsstrahlung.
In our model, we have similar mono-photon events from
$ e^+ e^- \to \nu_{e,\tau} \nu_{e,\tau} \, \ga $ via the $s_1$-exchange.
The constraint on DM searches can be translated as 
\begin{align}
\frac{1}{\La^4_{DM}} \gtrsim \frac{\th^4}{16 \, m^4_1} \lee \vert \la_e \vert^4  +  2 \vert \la_e \la_\tau \vert^2  +   \vert \la_\tau \vert^4  \rii ,
\end{align}
where the coefficient $1/16$ on the right-hand side is to account for the fact only the right-handed $e^-$~(left-handed $e^+$) and left-handed $\nu$~(right-handed $\bar{\nu}$)
are involved, i.e., $\lee 1/2 \times 1/2 \rii^2 $, and $\La_{DM} \simeq 320$ GeV for very small DM masses~\cite{Fox:2011fx}.

In the limit of $\la_e \sim \la_\tau$, the constraint is reduced to 
\begin{align}
\frac{1}{\La^2_{DM}} \gtrsim  \frac{\th^2 \la^2}{2 \, m^2_1}, 
\end{align}
and from Eq.~\eqref{eq:s12_NSI}, it implies the maximum NSI is
\begin{align}
\vert \vep^{s_1} \vert &= \frac{1}{2 \sqrt{2} G_F} \lee \frac{ \la^2 \theta^2}{ 2\, m^2_1} \rii \lesssim 0.3 ,
\end{align}  
which is consistent with results in Refs~\cite{Berezhiani:2001rs,Wise:2014oea} based on $e^+ e^- \to \nu_{e,\tau} \nu_{e,\tau} \, \ga$ analysis in the context of NSI.
Note that the mono-photon bound on NSI is unavoidable in the model since it is the same interactions
that contribute to both NSI and the mono-photon signals.

The bound derived above is actually more stringent than needed since  all the relevant processes in question are $t$-channel ones, and so one has, for the propagator,  
$\vert 1/((p_e - p_\nu)^2 -m^2_1 ) \vert \simeq  \vert 1/( 2 p_e \cdot p_\nu -m^2_1 )  \vert \lesssim 1/m^2_1 $, where $p_{e~(\nu)}$ is the
four momentum of the initial electron~(final neutrino). The contribution to mono-photons from $s_2$ also exists but is highly suppressed 
due to $1 /m^2_2 \lesssim 10^{-3} \, \th^2/m^2_1 $ necessitated for large NSI as explained in Eq.~\eqref{eq:FT_le}.

\item $\tau \to 3 e$ limit.\\
$\eta^0$ will induce $\tau^- \to  e^+ e^- e^-$ decay\footnote{Note that $\phi^\pm$ will {\it not} induce $\tau^- \to e^- \nu_{\tau}  \overline{\nu_e}$
at tree level since $\eta$ only couples to $e_R$ but not $\tau_R$.
Therefore, we will not consider bounds from the $\tau^- \to e^- \nu  \overline{\nu}$ branching ratio measurements.} and the decay width normalized to the $W$-mediated $\tau^- \to \mu^- \nu_\tau \bar{\nu}_\mu$ is constrained by
null $\tau \to 3e$ results from Belle Collaboration~\cite{Hayasaka:2010np},
\begin{align} 
\frac{\Ga_{\tau \to 3 e}}{\Ga_{\tau \to \mu \nu \bar{\nu}}}= \left| \frac{\la_e \la_\tau}{4 \sqrt{2}\, m^2_2 \, G_F} \right|^2 \lesssim \frac{2.7 \times 10^{-8}}{0.17} ,
\end{align}
which can be rewritten as 
\begin{align} 
 \frac{\sqrt{\la_e \la_\tau}}{m_2}  \lesssim \frac{0.16}{\text{TeV}},
\end{align}
and is stronger than the LEP constraints on $e^+ e^- \to \ell^+ \ell^-$. Note that one can similarly impose the bound from $\mu \to 3 e$ measurements
if $\la_{\mu}$ is switched on. Due to the fact $\text{Br}(\mu \to 3e) < 1.0 \times 10^{-12}$, one will have
$\sqrt{\la_e \la_\mu} / m_2  \lesssim 8.12 \times 10^{-3} / \text{TeV}$.
It implies that in order to achieve a sizable $\epsilon_{e \mu}$, the fine-tuning between $\mu^2_\phi$ and $\ka^2 v^2/ \mu^2_\eta$ mentioned in Eq.~\eqref{eq:FT_le}
has to be at the level of $10^{-5}$.   

 As mentioned above, sizable NSI induced via the $\eta-\phi$ mixing fall into 
the category of $d-8$ operators for which there is not always direct correlation between 
NSI and  CLFV interactions:
$\tau^- \to  e^+ e^- e^-$ does {\it not} receives the same enhancement from the $s_1$-exchange as NSI. 

\item $\tau \to e \ga$ bound.\\
$\eta^0$ will also radiatively induce $\tau \to e \ga$ and the process actually stems from the process $\tau \to 3 e$ by closing the $e^+$ and $e^-$ lines with a photon insertion.
Therefore, $\tau \to e \ga$ is suppressed by two powers of the electric coupling constant as well as a loop factor, which amount to $10^{-3}$ or so compared to $\tau \to 3 e$.
Given that the experimental constrains on these two processes are similar, $\text{Br}(\tau \to e \ga) \lesssim 3.3 \times 10^{-8}$~\cite{Aubert:2009ag} versus
$\text{Br}(\tau \to 3e ) \lesssim 2.7 \times 10^{-8}$~\cite{Hayasaka:2010np}, we will not include the $\tau \to e \ga$ limit here.

\end{itemize}

\begin{figure}[htbp!]
\centering
\includegraphics[clip,width=0.45\linewidth]{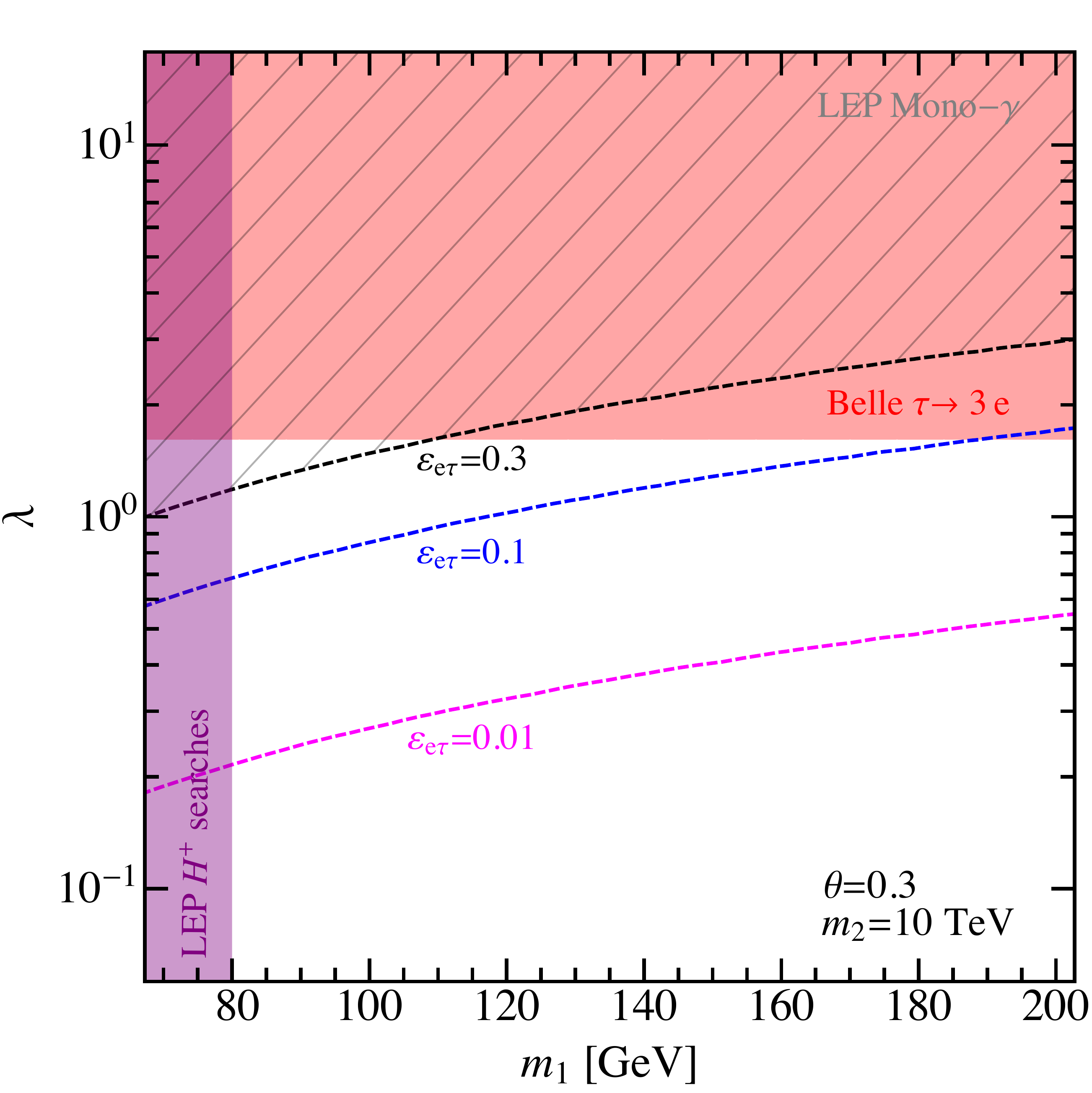}
\caption{ The summary plot of the constraints where we assume $\la_e=\la_\tau$, $\th=0.3$ and $m_2=10$ TeV.
The purple, red and crosshatched areas are excluded by the LEP charged Higgs  searches, Belle $\tau \to 3 e$ limit and LEP mono-photon bound, respectively.
To realize $\vep_{e\tau}=0.3$, $m_1$ has to range from 80 to 105 GeV. }
\label{fig:con_sum}  
\end{figure}

The constraints are summarized in Fig.~\ref{fig:con_sum}, where we choose $\th =0.3$ and $m_2=10$ TeV.
The purple area is excluded by the LEP searches on the charged Higgs, the light red area is eliminated by the Belle $\tau \to 3 e$
bound which are more stringent than the LEP bounds on $e^+ e^- \to \ell^+ \ell^-$, while the crosshatched region will be disfavored by the
LEP mono-photon searches. To achieve sizable NSI of $0.3$, $m_1$ is constrained to be between 80 and 105 GeV.   

Finally, we comment on the constraint from the electron magnetic dipole moment
and implications on the IceCube experiment.
At one-loop level, the electron anomalous magnetic moment $(g - 2)$ receives an additional radiative contribution from loops of $\phi^-$ and $\nu_e$.
The contribution can be estimated as:
\begin{align}
\De a_e \equiv \frac{g-2}{2} \sim \frac{e \, \la^2_e \, \th^2}{16 \pi^2} \frac{m^2_e}{ m^2_\phi} \sim 10^{-14},
\end{align}
for the region of interest in Fig.~\ref{fig:con_sum}. It is much smaller than the difference between the experiment result and the SM prediction:
 $a^{\text{exp}}_{e} - a^{\text{SM}}_{e} = - (1.06 \pm 0.82) \times 10^{-12}$~\cite{Hanneke:2008tm, Hanneke:2010au,Aoyama:2012wj,Endo:2012hp}.
Therefore, the new contribution to $ \Delta a_e$ is negligible.

As pointed out in Ref.~\cite{Dutta:2015dka}, the resonance enhancement with a 
single scalar leptoquark
can be used to increase the very high energy shower event rates at the IceCube.
In our model, high energy neutrinos can interact with electrons and
produce $\phi$ or $\eta$, which later decays into neutrinos and charged leptons.
To have $\phi$ or $\eta$ on-shell, one must have $\sqrt{2 m_e E_\nu} \gtrsim m_{\phi,\eta}$,
requiring $E_\nu \sim 10$ and $10^5$ PeV for $m_\phi \sim 100$ GeV and $m_\eta \sim 10$ TeV.
The flux of such high energy cosmic neutrinos is then highly suppressed. 
Moreover, the relevant coupling for the $\phi$-exchange has to be of $\mathcal{O}(1)$~\cite{Dutta:2015dka} so that the new contribution is comparable with that of the SM.
Nonetheless, the coupling in this model is simply $\lambda^4 \th^4$, that is much  smaller than the unity for regions of interest.
As a result, one can not account for PeV events at the IceCube with the resonance enhancement 
of $\phi$ or $\eta$.

\section{NSI Oscillations at DUNE \label{section:analysis}}

A NC-like NSI interaction can be parametrized as four-fermion effective 
operators of the form:

\begin{equation}\label{eq:NSIlag}
\mathcal{L}^{NSI}_{NC}=-2\sqrt{2}G_{F} \sum_{f} \! 
\eps^{f,P}_{\alpha\beta} \!
      \left[ \bar{\nu}_\alpha \gamma^{\rho}L \nu_\beta \right] 
\!\!
      \left[ \bar{f} \gamma_{\rho} P f \right]
\end{equation}
where $G_{F}$ is the Fermi constant, $\eps^{f, P}_{\alpha\beta}$ are the 
NSI dimensionless couplings whose absolute value represent the relative NSI 
strength, $P=(L,R)$ is the chiral projector, and $f$ is the SM fermion of the first family: 
$e$, $u$, and $d$. 

The NSI effective interactions modify the effective matter potential that accounts for 
the neutrino--mater interactions. Therefore, there is a dependency on the fermion density 
in the medium. For long baselines below $2000\,\text{km}$, one can 
assume the matter density is constant simplifying the expression for the Hamiltonian in 
presence of the NSI, which can be written as:
\begin{equation}\label{eq:Hint}
H_{\text{int}}=V
\left(\begin{array}{ccc}
 1+\eps_{ee} & \eps_{e\mu} & \eps_{e \tau} \\
 \eps_{e\mu}^* & \eps_{\mu \mu} & \eps_{\mu \tau} \\
 \eps_{e \tau}^* & \eps_{\mu \tau}^* & \eps_{\tau \tau}
\end{array} \right)
\end{equation}
with $V=\sqrt{2}\,G_F\,N_e$, where $N_e$ is the electron density on Earth. Notice that 
the `$1$' in the interaction Hamiltonian corresponds to the SM neutrino--matter 
interactions. By adding the NSI coupling in the formalism, we have increased the number 
of real parameters by eight since, in the diagonalization, one of the diagonal parameters 
can be rephased out. It is worth to mention that long baseline neutrino oscillations are 
sensitive to a combination of NSI couplings defined in Eq.~(\ref{eq:NSIlag}):

\begin{equation}\label{eq:nsi-lag}
\varepsilon_{\alpha \beta}=\sum_{f=e,u,d}\left \langle \frac{Y_f}{Y_e} \right \rangle 
\varepsilon_{\alpha \beta}^f= \varepsilon_{\alpha \beta}^e+Y_u\varepsilon_{\alpha 
\beta}^u+Y_d\varepsilon_{\alpha \beta}^d
\end{equation}
where $Y$ is the abundance of each fermion in the medium. In Eq.~(\ref{eq:nsi-lag}) the 
effective NSI couplings are a weighted combination of the Lagrangian parameters. 

The vacuum neutrino oscillations are governed by the usual Hamiltonian 
\begin{equation}\label{eq:H_0}
H_0=\frac{1}{2E}\, \left[
U\,\text{diag}\{0,\Delta m_{21}^2,\Delta m_{31}^2\}\,U^\dagger \right]
\end{equation}
where $U$ is the lepton mixing matrix, $\Delta m^2_{k1}$ are the two measured mass 
squared differences, and $E$ is the energy of the incoming neutrino.  The total 
Hamiltonian describing neutrino oscillations in matter is the sum of Eq.~(\ref{eq:Hint}) and Eq.~(\ref{eq:H_0}).

In our model, we have NSI in the lepton sector only, i.e., 
$\vep^u=\vep^d=0$ and thus $\vep=\vep^e$, since the imposed $Z_2$
symmetry forbids Yukawa couplings of $\eta$ to quarks. To simplify the analysis, we set
$\la_e=\la_{\tau}$~\footnote{Since one of the diagonal NSI parameters is irrelevant in 
the diagonalization of the Hamiltonian in Eq.~(\ref{eq:Hint}), one can set $\eps_{\mu 
\mu}$ equal to zero which implies $\la_\mu=0$. However, $\la_\mu \ne 0$ also induces 
two off-diagonal terms in addition to diagonal one. The resulting off-diagonal terms in 
principle affect the oscillation NSI analysis although the effect is small. Therefore, in 
our analysis we have assumed $\la_\mu=0$ without significantly affecting the 
CP degeneracy mentioned above.}, where $\la_\al$ is the Yukawa coupling of $\eta$ defined in 
Eq.~\eqref{eq:eta_Yuk}.  From Eq.~\eqref{eq:eps_al_be}, we have the following relations:

\begin{align}\label{eq:corr}
\vep_{e e} &=\vep_{\tau \tau} \equiv  - \vert \vep \vert \nn \\
\vep_{e \tau} &= \vert \vep \vert  \exp{(i\,\phi)} ,
\end{align}
which are similar to those of a recent work~{\cite{Farzan:2016wym}}, that features a light gauge boson $Z^\prime$ corresponding to the $U(1)_B$ or $U(1)_{B-L}$ gauge symmetry and can also generate large NSI, including
$\vep_{e\tau}$.

For the numerical analysis we have used the GLoBES 
library~\cite{Huber:2004ka,Huber:2007ji} and the NSI tool (prepared for the study in 
Ref.~\cite{Kopp:2007ne}) with the official implementation of the DUNE experiment from 
Ref.~\cite{Alion:2016uaj}. In the analysis we have included the full DUNE implementation, 
i.e. the four oscillation channels for (anti-)neutrino appearance and disappearance 
running $3.5$ years in each mode with the 
optimized neutrino beam. Also, we included the  effect of the systematical errors in our 
analysis. Finally, as `true' parameters, we used the best-fit values for the standard 
oscillation parameters from Ref.~\cite{Forero:2014bxa} except for the reactor mixing 
angle, whose value was fixed to the Daya Bay result from Ref.~\cite{An:2015rpe}. The 
atmospheric mixing angle is assumed maximal but large errors on the 
atmospheric parameters (with the current precision) were implemented as penalties in the 
$\chi^2$ statistical analysis. Our analysis is based on the normal 
neutrino mass hierarchy~(NH) and we commented on the relevant differences in the case of the 
inverted mass hierarchy~(IH) at the end.

\begin{figure*}[t!]
\includegraphics[width=0.49\columnwidth]{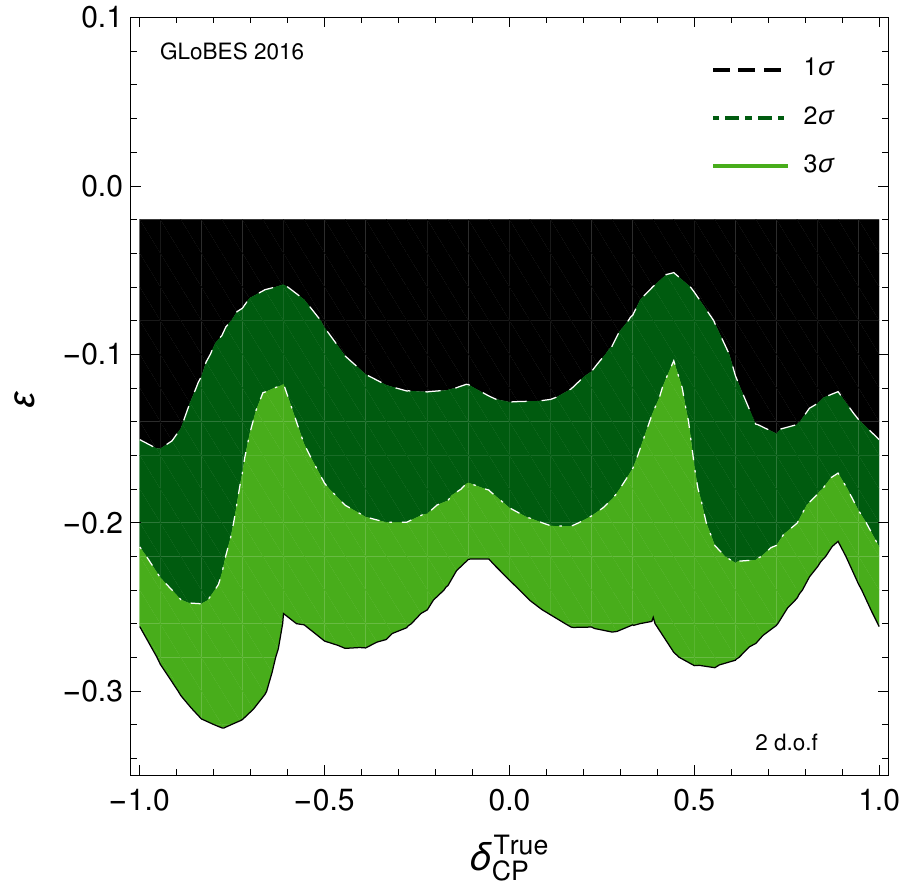}
\includegraphics[width=0.49\columnwidth]{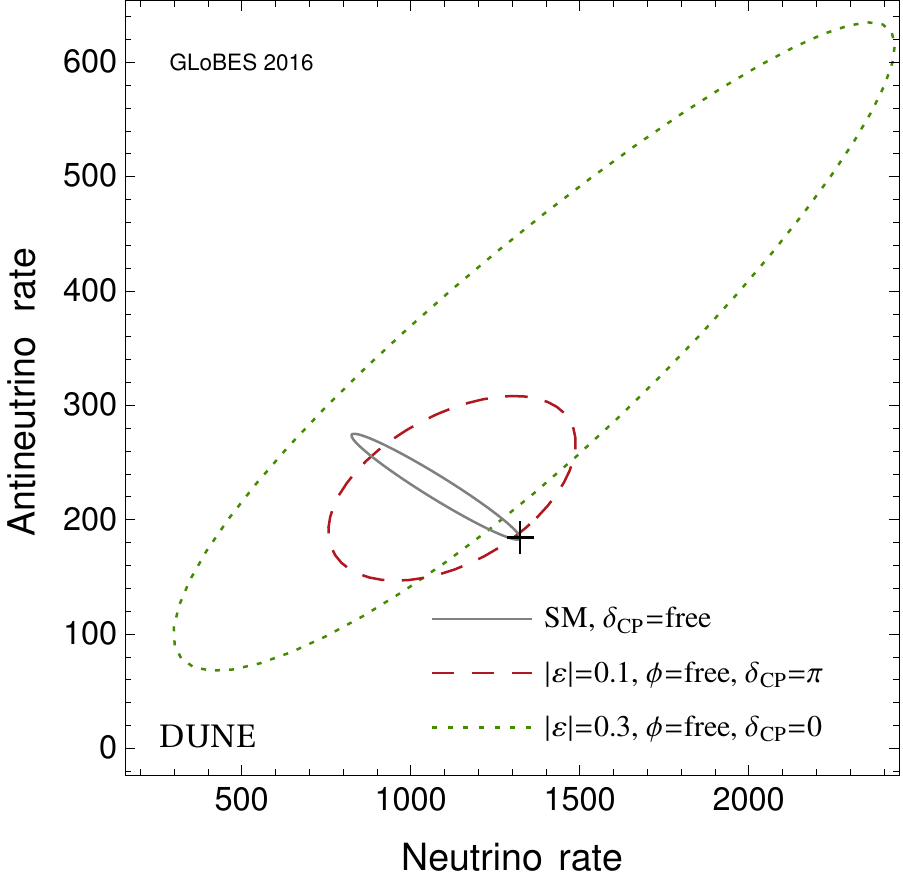}
 \caption{The left panel corresponds to the DUNE sensitivity to the NSI parameter space 
defined in Eq.~(\ref{eq:corr}). Most of the parameters not shown in the plot were  
marginalized over. See text for details. In the right panel, we show the bi-rate plots 
that identifies the parameter degeneracies. The solid line corresponds to the case 
with SM interactions and for all Dirac CP phase values. The dashed and dotted curves 
correspond to the NSI case, for all possible NSI phase $\phi$ values, and for CP 
conserving values of the Standard CP phase. The SM case with $\delta_{\text{CP}}=-\pi/2$
denoted by the cross is also shown including the statistical uncertainty as a reference. In 
this case, the standard oscillation parameters not shown were fixed to their best-fit values.}
\label{fig:bi-rate}
\end{figure*}

Initially, we have extracted a constraint on the NSI couplings in our simplified setup by 
assuming only standard oscillation parameters as the `true' parameters and by testing 
the NSI couplings. The results are shown in the left panel of Fig.~\ref{fig:bi-rate} 
where the dependency on the `true' Dirac CP phase value is also shown. All the parameters 
not shown in the plot have been marginalized over except for the solar oscillation 
parameters and the reactor mixing angle that were fixed to their best-fit 
values~\footnote{The central values and uncertainties for the oscillation parameters ($\theta_{ij}$, $\Delta 
m_{k1}^2$), that are marginalized over, are 
obtained from the global fit analysis~\cite{Forero:2014bxa} assuming standard interactions 
only, i.e., in the absence of NSI.}. We have obtained the allowed interval $\eps \in [-0.16,0]$ 
at the $90\%$ confidence level for 
$1\,\text{d.o.f.}$. This limit can not be directly compared with with existing works in 
Refs.~\cite{Coloma:2015kiu,deGouvea:2015ndi}
due to the correlations in 
Eq.~(\ref{eq:corr}) from our model. Notice that, in our model, only the $\eps^e$ 
couplings are predicted and therefore NSI constraints from neutrino--electron scattering 
also apply. However, the bound extracted from DUNE simulated data is compatible with the 
scattering NSI bounds in 
Refs.~\cite{Forero:2011zz,Barranco:2007ej,Khan:2016uon} by identifying $\eps^e=\eps^{e 
R}+\eps^{e L}$.

In order to evidence the parameter degeneracies after the inclusion of the NSI couplings, 
we have made use of the total signal rates shown in the a bi-rate plot in the right 
panel of Fig.~\ref{fig:bi-rate}. In the same spirit of Ref.~\cite{Forero:2016cmb}, the 
`true' Dirac CP phase values were assumed to be CP conserving to explore the 
possibility that the new phase, coming from the NSI, could mimic the effect of the standard 
Dirac CP phase -- what we call the `confusion'. For comparison, the standard oscillation 
case is also shown with a benchmark point $\delta_{CP}=-\pi/2$ including the statistical errors. This point is one of the probable 
values within the allowed range of the Dirac CP phase determined by the T2K and 
NOvA~\cite{Abe:2015awa,Adamson:2016tbq} measurements after including the reactor mixing 
angle determined at reactors. In the case the value 
$\delta_{CP}=-\pi/2$ were measured at DUNE a potential of `confusion' might arise after 
the inclusion of the NSI. In particular, this is more evident for the case of $|\eps|=0.1$ 
with $\delta_{CP}=\pi$ since the maximum neutrino rates are comparable with the 
SM prediction with $\delta_{CP}=-\pi/2$. Notice however that if one includes also the 
statistical errors for the NSI and SM ellipses, there is an ample room for the confusion to 
happen. In other words, considering that the current preferred values span the complete 
negative region of the  parameter $\delta_{CP}/\pi$, there is a potential of confusion 
for values of the CP violating phase within the interval $[-1,0]$. 

\begin{figure*}[t!]
\includegraphics[width=0.49\columnwidth]{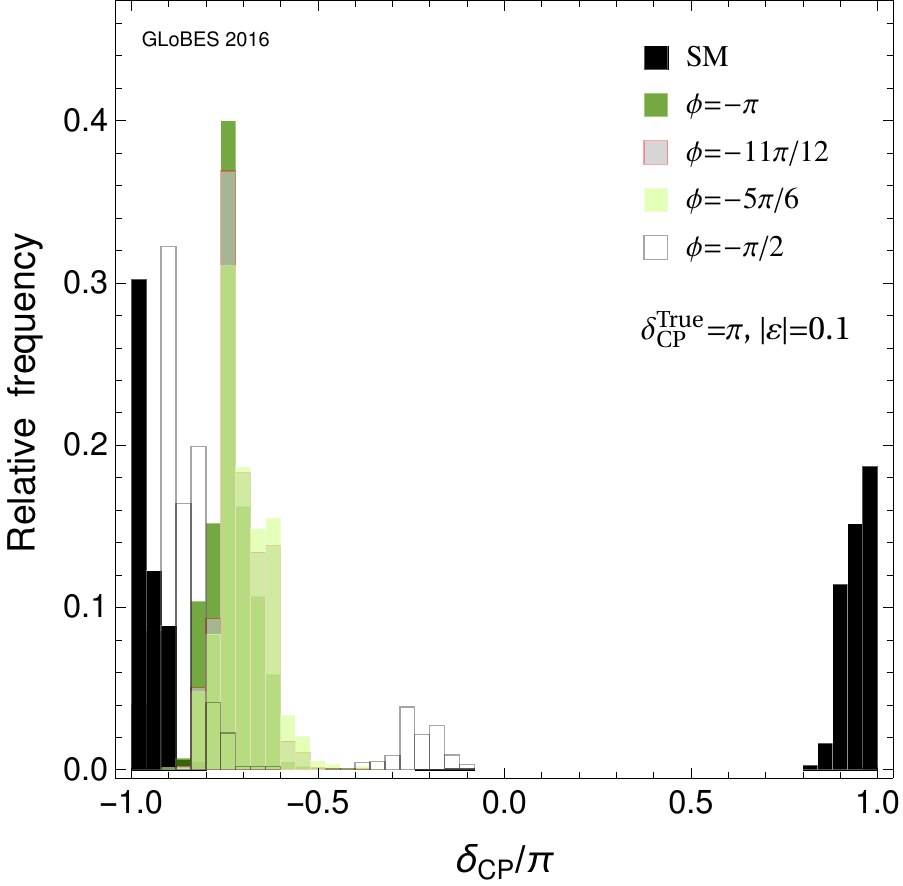}%
\includegraphics[width=0.49\columnwidth]{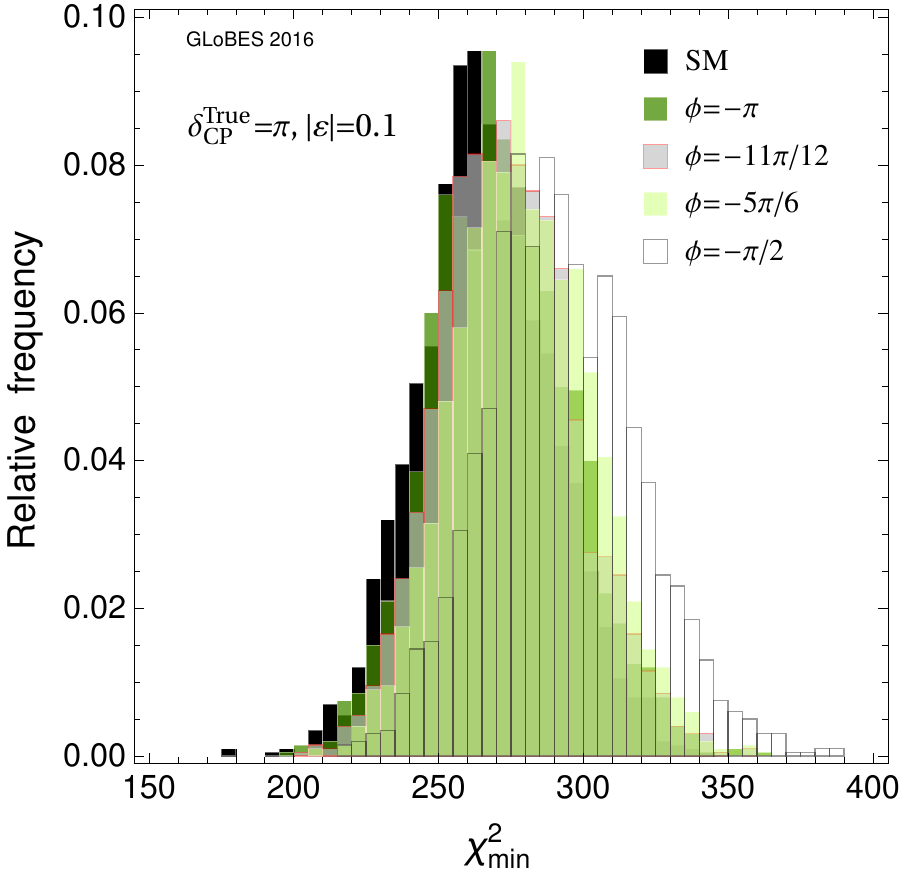}
 \caption{Results assuming $\delta_{\text{CP}}^{\text{True}}=\pi$. In
   the left panel, the best-fit distributions for the Dirac CP phase are shown for both
   the SM and NSI cases. We have fixed the NSI magnitude to the value 
$|\varepsilon|=0.1$ (see also Eq.(\ref{eq:corr})) and for the different NSI phases $\phi$ 
showed in the plot. In the right panel the minimum $\chi^2$ distributions are shown for 
the SM and NSI cases showed in the left panel. All not shown parameters were marginalized 
over,
   see text for details of the analysis.}
\label{fig:histogramCP180}
\end{figure*}

We now are in a position to quantify the degree of `confusion' in the establishment of CP 
violation in the lepton sector in DUNE.  The magnitude of the NSI couplings regarded as a `true' parameter
is fixed to $|\eps|=0.1$ and the SM CP phase is chosen to be 
$\delta^{True}_{CP}=\pi$. Statistical fluctuations in the `true' rates are included 
and we test the standard oscillation rates. In the left panel of 
Fig.~\ref{fig:histogramCP180}, we show the distribution of the best-fit value of $\delta_{CP}$ for
different values of the new phase $\phi$. For comparison, we also 
display the case without NSI, which appears distributed 
around $\delta^{True}_{CP}=\pm \pi$ as expected.
Given the chosen values of the NSI phase $\phi$~($\phi=-\pi$, $-11\pi/12$, $-5 \pi/6$, $-\pi/2$),
the histograms cover the interval $[-\pi,\pi/2]$. 
Except for the case of $\phi=-\pi/2$, the NSI histograms are centered around the particular 
value of $-3\pi/4$ ($-135^\circ$). Even though none of the NSI histograms is centered
around the reference value `$-\pi/2$' (the current best-fit value), there is however some potential of `confusion'.
After all, with the current data, the allowed range of $\delta_{CP}/\pi$ is
$[-0.996,-0.124]$ at the $90\%$ confidence level for the normal hierarchy~\cite{t2k-ICHEP:16}. 
Thus, if DUNE measures a value of the CP violating phase close to 
$-3\pi/4$~(away from the current best-fit 
value), the degeneracy will persist. Otherwise, DUNE might break the degeneracy, 
strongly depending on its precision on the CP violating phase measurement.

The minimum $\chi^2$ distributions for each of standard and NSI cases in the 
left panel of Fig.~\ref{fig:histogramCP180} are shown in the right panel of the same 
figure. Except for the case of $\phi=-\pi/2$, the histograms of the NSI and the 
standard cases are centered around $\chi^2_{min}\sim 260$, which is compatible with the number of bins
minus that of the fitted parameters, and is within a deviation of less 
than ten $\chi^2_{min}$ units. This evidences the possibility that DUNE might not have 
the ability to distinguish the origin of the CP violation if the measured CP phase 
happens to be around $-3\pi/4$.

Finally, in the case of the IH, the constraint on $\vert \varepsilon \vert$ is similar to 
that of the NH case, shown in the left panel of Fig.~\ref{fig:bi-rate} and 
it is even stronger for certain values of 
$\delta_{\text{CP}}^\text{true}$. The parameter degeneracy shown in the right panel of 
Fig.~\ref{fig:bi-rate} for NH is also present in the IH case for different NSI parameters, in 
particular for $|\vep|=0.12$ with $\delta_{CP}=0$. This is due to the fact that for IH in DUNE, 
with $\delta_{CP}=-\pi/2$, lower neutrino rates and higher antineutrino rates are expected in 
comparison to the NH case. With $|\vep|=0.12$ and 
$\phi=\pi/3$, the histogram of relative frequency~(the left panel 
of Fig.~\ref{fig:histogramCP180} for NH) will be centered around $\delta_{CP}=\pi/4$ with 
$\delta^{\text{True}}_{\text{CP}}=0$ in the case of IH. Clearly, the value is 
in tension with the current preferred one $\delta_{CP} \sim -\pi/2$~\cite{t2k-ICHEP:16} and therefore it is likely that DUNE will break the 
degeneracy. Notice that we here have fixed $\delta^{\text{True}}_{\text{CP}}=0$
but one can have mixed sources of CP violation from both the SM and NSI. 
In contrast, in the NH case even for a CP conserving 
value $\delta^{\text{True}}_{\text{CP}}=\pi$,
there exists the degeneracy which may not be resolved by DUNE, 
as demonstrated in Fig.~\ref{fig:histogramCP180}.

\section{Conclusions \label{section:conclusions}}

We come up with a novel way to achieve sizable NSI of order $\mathcal{O}(0.3)$ at the cost of
fine-tuning, which is required to be at the level of $10^{-3}$. 
In addition to the SM particles, the extra $SU(2)$ doublet and charged singlet scalars, denoted by $\eta$ and $\phi$
respectively, are introduced as well as the $Z_2$ symmetry, under which $\eta$, $\phi$ and the right-handed electron~$e_R$ are odd. The charged
component of $\eta$ mixes with $\phi$ via the dimensionful coupling $\ka$ to the Higgs boson, $\ka \phi \eta H$ with $\lan H \ran =v$.
Note that the $Z_2$ symmetry is explicitly broken by the electron Yukawa coupling and it is plausible that the smallness of the coupling is protected by the $Z_2$ symmetry. 

If there exists the mass hierarchy, $m^2_\eta \gg \ka v \gg m^2_\phi$ , the mass of the light mass eigenstate $s_1$ can be treated as an uncorrelated
parameter from the $\phi - \eta$ mixing angle at the price of fine-tuning.
As a result, one can have a very small mass of $s_1$, $m_1$, but a relatively large mixing angle $\th$.
Note that without fine-tuning one has $\th \sim m_1/ m_2 $, where $m_2$ is the mass of the heavy eigenstate $s_2$.

An analogy can be drawn between this model and hybrid models of the Type-I plus Type-II seesaw mechanism,
where light neutrino masses similarly receive two contributions from
the mixing with heavy right-handed neutrinos and from the VEV of the $SU(2)_L$ triplet scalar. The light-heavy neutrino mixing angle  is merely determined by the heavy neutrino mass and the Yukawa coupling but is not related to the triplet VEV.

NSI can be generated by coupling $\eta$ to $e_R$ and the $SU(2)_L$ lepton doublets $L_\al$~($\al = e, \tau$), i.e.,
$\la_\al \overline{L_\al} \, \eta \, e_R$ which obeys the $Z_2$ symmetry.  The new Yukawa coupling will give rise to NSI via the $\eta^\pm$-exchange. $\la_\mu$ is not considered here since it has a little impact on the `confusion' we look for.
In light of the $\phi - \eta$ mixing, NSI from $s_1$ is $\la_\al \la_\beta \th^2 / m^2_1$, which can be sizable if $\th^2/m^2_1 \gg 1/m^2_2$. Taking into account various experimental bounds such as the LEP measurements on the $e^+ e^- \to \ell^+ \ell^-$ cross-section,
upper limits on $\tau \to 3 e$, $\tau \to e \ga$ branching ratios, $\la_{e,\tau}$ are constrained to less than 0.16 for TeV $s_2$, while
the LEP searches on the charged Higgs
demand $m_1$ to be greater than 80 GeV.
All in all, one needs $m^2_1/m^2_2 \sim 10^{-3} \, \th^2$ such that $\vep_{e \tau}$ can be as large as 0.3 with $m_1 \sim 100$ GeV,
given $\th \sim 0.3$ and $m_2 \gtrsim 10$ TeV.
The inevitable upper bound on $\vep_{e \tau}$ comes from the LEP mono-photon searches since in our model both NSI
and mono-photon signals result from exactly the same interactions.

The extra particles $\eta$ and $\phi$ in the model are within the reach of future experiments. First, the LEP mono-photon search has limited $\vep_{e \tau}$ to be smaller than $0.3$. Future electron colliders such as ILC~\cite{BrauJames:2007aa,Djouadi:2007ik} or FCC-ee~(formerly known as TLEP,)~\cite{Blondel:2011fua,Gomez-Ceballos:2013zzn}
can significantly improve the mono-photon bound or spot the signal, which is an indirect evident of the charged $\phi$. Second, the accessible branching fractions for $\tau \to 3 e$ at the superKEKB/Belle II will reach the level of $\mathcal{O}(10^{-10})$~\cite{Abe:2010gxa,Hayasaka:2013dsa},
discovery of $\tau \to 3e$ or $\tau \to e \ga$ will implicitly indicate the presence of the neutral component $\eta^0$. Third, the direct detection of $e^+ e^- \to e^+ \tau^-$ or $e^- \tau^+$ at high-luminosity ILC and FCC-ee will be a
smoking gun for the $\eta^0$ existence. 

We have also discussed the phenomenological implications from the induced NSI.
One of the main objectives of the future neutrino program is to establish if 
there is CP violation in the lepton sector with the help of current and future 
facilities. DUNE is one of the future facilities that will shed light on the current 
unknowns in the three neutrino framework and in particular on the determination of the 
CP violating phase. In this work we have determined DUNE 
sensitivity to the generated NSI by taking a simple limit of $\la_e=\la_\tau$, which results in
the correlations  in Eq.~(\ref{eq:corr}). We have extracted the bound $\eps \in 
[-0.16,0]$ at the $90\%$ confidence level Since DUNE is sensitive to an NSI at the $\sim 10\%$ level, 
we also studied the NSI impact on the determination of the CP violating phase. To this 
purpose, we have exploited the parameter degeneracies that arise due to the new 
parameters coming form the NSI. One of the main consequences is the possible `confusion' 
in terms of the source of the CP violation. We have study the degree of `confusion' at 
DUNE experiment by setting the Dirac CP phase to a CP conserving value and allowing the 
$\eps_{e \tau}$ NSI phase to generate the observed CP violation. We have found that if 
DUNE measures a phase close to $-3\pi/4~(-135^\circ)$ instead of $-\pi/2$, DUNE will not be 
able to determine the origin of the measured CP violation. Otherwise, if DUNE 
measures a CP phase different from $-3\pi/4$ with a precision better than 
$\sim 10^\circ$ then it will be able to break the standard and NSI CP degeneracy studied 
here.

Finally, we would like to point out that by having $\eta$ couple to quarks, one also get 
large NSI from quark-neutrino interactions. It is possible to realize the ``dark-side'' 
solution for solar neutrinos
proposed in Ref~\cite{Miranda:2004nb,Escrihuela:2009up}
as an alternative to the standard LMA solution based on the Mikheev-Smirnov-Wolfenstein mechanism~\cite{Wolfenstein:1977ue,Mikheev:1986gs}.
Bounds on NSI from LHC mono-jet searches~\cite{Friedland:2011za}, however, will come into play in this case. 
Some models~\cite{Farzan:2015doa,Farzan:2015hkd,Farzan:2016wym}
have recently been proposed to realize such large NSI. 

\section*{Acknowledgments}
The authors would like to thank Andr\'e de Gouv\^ea and Sofiane M. Boucenna for helpful 
discussions, and thank Joachim Brod and Andr\'e de Gouv\^ea for useful comments on the 
draft. WCH is grateful for  the hospitality of Northwestern University HEP group where 
the project was initiated. WCH is supported by DGF Grant No. PA 803/10-1. DVF thanks 
the URA fellowship that allowed him to visit the theory division at Fermilab where this 
project was initiated. DVF has been supported by the U.S. Department of Energy under the
DE-SC0013632 and DE-SC0009973 contracts.



\bibliographystyle{h-physrev}
\bibliography{NSIetau,NSIetauAddRefs}

\end{document}